\documentstyle[preprint,aps,epsfig]{revtex}
\def\beq{\begin{equation}}
\def\be{\begin{eqnarray}}
\def\eeq{\end{equation}}
\def\ee{\end{eqnarray}}
\def\ci{\cite}
\def\bi{\bibitem}

\begin{document}

\draft

\title{$Q^2$-dependence of backward pion multiplicity
in neutrino-nucleus interactions}

\author
{ O. Benhar$^{1}$,  S. Fantoni${^2}$, G.I. Lykasov${^3}$
\cite{byline}, U. Sukhatme $^4$ }

\address
{ $^1$ INFN, Sezione Roma 1, I-00185 Rome, Italy \\
$^2$ International School for Advanced Studies (SISSA), I-34014 Trieste, Italy\\
$^3$ Joint Institute for Nuclear Research, Dubna 141980, Moscow
Region, Russia \\ 
$^4$ Dept. of Physics, University of Illinois,  Chicago, IL 60607, USA}

\date{\today}

\maketitle


\begin{abstract}

The production of pions emitted backward in inelastic 
neutrino-nucleus interactions 
is analyzed within the impulse approximation in the framework
of the dual parton model. We focus on the 
$Q^2$-dependence of the multiplicity of negative pions, 
normalized to the total
cross section of the reaction $\nu + A \rightarrow \mu + X$.
The inclusion of planar (one-Reggeon 
exchange) and cylindrical (one-Pomeron exchange) graphs
leads to a multiplicity that decreases as $Q^2$ increases, in 
agreement with recent measurements carried out at CERN by 
the NOMAD collaboration. A realistic treatment of the 
high momentum tail of the nucleon distribution 
in a nucleus also allows for a satisfactory description of the 
semi-inclusive spectrum of backward pions.

\end{abstract}


 


Over the past several years, backward hadron production in inelastic 
lepton-nucleus scattering has been extensively investigated, both
experimentally \ci{backprot,backpion,NOMAD} and theoretically 
\ci{cls,diper,bfl1,bfl2,bfl4}, with the aim of pinning down 
the dominant reaction mechanism and extracting information on the underlying 
dynamics.

The analysis carried out on refs.\ci{bfl1,bfl2} focused on 
nuclear effects in $e + A \rightarrow e^\prime + p + X$ reaction, 
while in ref.\ci{bfl4}, hereafter referred to as I, we studied
the spectrum of pions emitted in the process
$\nu + A \rightarrow \mu + \pi + X$.
The results in I show that the main contribution to this spectrum 
comes from scattering off nucleons carrying high momentum.
The calculations were performed within the impulse 
approximation (IA) scheme, using the Quark Gluon String Model 
(QGSM) \ci{kaid1,kaid2} to describe the interaction vertex. This
model is similar to the Dual Parton Model (DPM) developed in 
refs.\ci{dpm1,dpm2}, in that both approaches are aimed at implementing 
the dual topological unitarization scheme. 
The relativistic invariant semi-inclusive spectrum of pions emitted 
strictly backward was obtained including only
the cylindrical (i.e. one-Pomeron exchange) graph, while the contribution 
of the planar (i.e. one-Reggeon exchange) diagram was neglected.

In this paper we continue the investigation of semi-inclusive pion 
production, started in I. We extend our analysis to pions emitted 
in the whole backward hemisphere, with respect to the 
direction of the incoming beam, and use QGSM to consistently 
include both the cylindrical and planar graphs in the calculation 
of the spectrum integrated over the transverse pion momentum.
Our results are compared to the new data from the NOMAD collaboration 
\ci{NOMAD}, which measured the spectrum of $\pi$ emitted in the backward
hemisphere in the reaction $\nu + C \rightarrow \mu + \pi + X$ and the
$Q^2$-dependence of the pion multiplicity, $\langle n_\pi \rangle$, 
normalized to the total cross section of the reaction 
 $\nu + C \rightarrow \mu + X$. 

Our study focuses on the kinematical region of
large pion momentum ($p_\pi > 0.3$ GeV/c), where the effects
of final state interactions (FSI) leading to pion absorption associated 
with production of baryon resonances are expected to be small
\ci{fs2,al} and the IA can be safely used.

The relativistic invariant semi-inclusive spectrum of 
 pions produced in the process 
$\ell + A \rightarrow \ell^\prime + \pi + X$, in which the incoming lepton 
is scattered with energy $E^\prime$ into the solid angle $d\Omega$, is 
defined as
\beq
\rho_{\ell A \rightarrow {\ell^\prime} \pi X} \equiv E_\pi
\frac{ d\sigma}{d^3p_\pi d\Omega dE^\prime}\ ,
\label{def:spec}
\eeq
where $E_\pi$ and ${\bf p}_\pi$ are the total energy and three momentum of
the produced pion, respectively. 
Within the framework of IA the quantity appearing on the right hand side 
of the above equation 
can be rewritten in convolution form, in terms of the semi-inclusive 
spectra of pions produced in 
lepton-proton and lepton-neutron processes
$\rho_{\ell p \rightarrow {\ell^\prime} \pi X}$ and  
$\rho_{\ell n \rightarrow {\ell^\prime} \pi X}$, according to
 (see, e.g., refs. \ci{bfl1,bfl4,fs1})
\be
\nonumber
\rho_{\ell A \rightarrow {\ell^\prime} \pi X}(x,Q^2;z,p_{\pi t}) 
& = & \int_{x,z \leq y}\  dy\ d^2k_t\  f_A(y,k_t) 
\ \left[ \frac{Z}{A} \rho_{\ell p \rightarrow {\ell^\prime} \pi X}
\left( \frac{x}{y},Q^2;\frac{z}{y},p_{\pi t}-\frac{z}{y}k_t \right) \right. \\
& + & \left.
\frac{N}{A} \rho_{\ell n \rightarrow {\ell^\prime} \pi X}
\left( \frac{x}{y},Q^2;\frac{z}{y},p_{\pi t}-\frac{z}{y}k_t \right) \right],
\label{def:lasp}
\ee
where $Z$ and $N$ are the number of protons and neutrons, $A=Z+N$ and
$Q^2=-q^2$, $q$ being the four-momentum tranferred by the lepton.
The relativistic invariant variables $x$ and $z$ are defined as
\ci{backpion,NOMAD}
\beq
x = \frac{M_A}{m}\ \frac{Q^2}{2(P_A \cdot p_\nu)}\ \ \ \ \ , 
 \ \ \ \ \ z = \frac{M_A}{m}\ \frac{(p_\pi \cdot p_\nu)}{(P_A\cdot p_\nu)}\ ,
\label{def:xz}
\eeq
where $P_A$ and $p_\nu$ are the four-momenta of the nucleus and the initial 
neutrino, respectively, and $M_A$ and $m$ are the nucleus
and nucleon masses. The above variables are best suited to analyze the data
of ref.\ci{NOMAD}, in which the backward direction is defined with respect to the 
beam direction. Note that they do not coincide with $x$ and
$z$ employed in I, whose definition can be recovered by replacing the 
four-momentum of the incoming neutrino, $p_\nu$, with the four momentum 
transfer $q$ in eq.(\ref{def:xz}). 

The nucleon distribution function
$f_A(y,k_t)$ can be written \ci{bfl1}
\beq
f_A(y,k_t) = \int dk_0\ dk_z\ S(k)\ y \delta\left(y-\frac{M_A}{m}
\frac{(k\cdot p_\nu)}{(P_A\cdot p_\nu)}\right)
\label{def:fa}
\eeq
where $S(k)$ is the relativistic invariant function describing the
nuclear vertex with an outgoing virtual nucleon.
The function $f_A(y)$ resulting from the 
transverse momentum integration of $f_A(y,k_t)$ can be obtained 
by approximating $S(k)$ with the nonrelativistic
spectral function $P({\bf k},E)$ yielding the probability of finding
a nucleon with momentum ${\bf k}$ and removal energy $E=m-k_0$
in the target nucleus \ci{bf1}. 
However, due to the limited range of momentum and
removal energy covered by nonrelativistic calculations of
$P({\bf k},E)$ (typically $|{\bf k}|<k_{max}\sim 0.7-0.8$  GeV/c
and $(m-k_0) < 0.6$ GeV. (see, e.g., ref. \ci{bf1}), this procedure
can only be used in the region $y < y_0 \sim 1.7-1.85$. An
alternative approach to obtain $f_A(y)$ at larger $y$, based
on the calculation of the overlap of the relativistic invariant
phase-space available to quarks belonging to strongly correlated
nucleons, has been proposed in ref.\ci{bfl1}. A similar procedure has 
been also used in ref.\ci{fs2} to obtain the quark distribution
in deuteron at large $y$.

The analysis of backward pion production requires
the full nucleon distribution function given by eq.(\ref{def:fa}). 
We assume that it can be written in the factorized form
\beq
f_A(y,k_t)=f_A(y)g_A(k_t)\ ,
\label{def:faykt}
\eeq
with the function $g_A$ chosen of the Gaussian form
\beq
g_A(k_t)=\frac{1}{\pi~ \langle k_t^2 \rangle }
{\rm e}^{-k_t^2/\langle k_t^2 \rangle}\ ,
\label{def:ga}
\eeq
where $\langle k_t^2 \rangle$ is the average value of the 
squared nucleon transverse momentum.

\begin{figure}
\vspace*{1.in}
\centerline
{\epsfig{figure=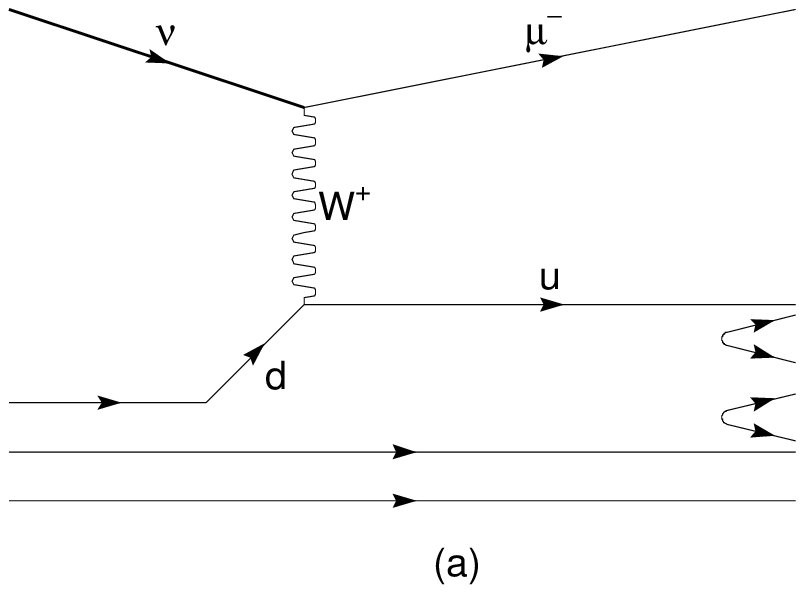
,angle=000,width=10.cm,height=8.0cm}}
\vspace*{.2in}
\centerline
{\epsfig{figure=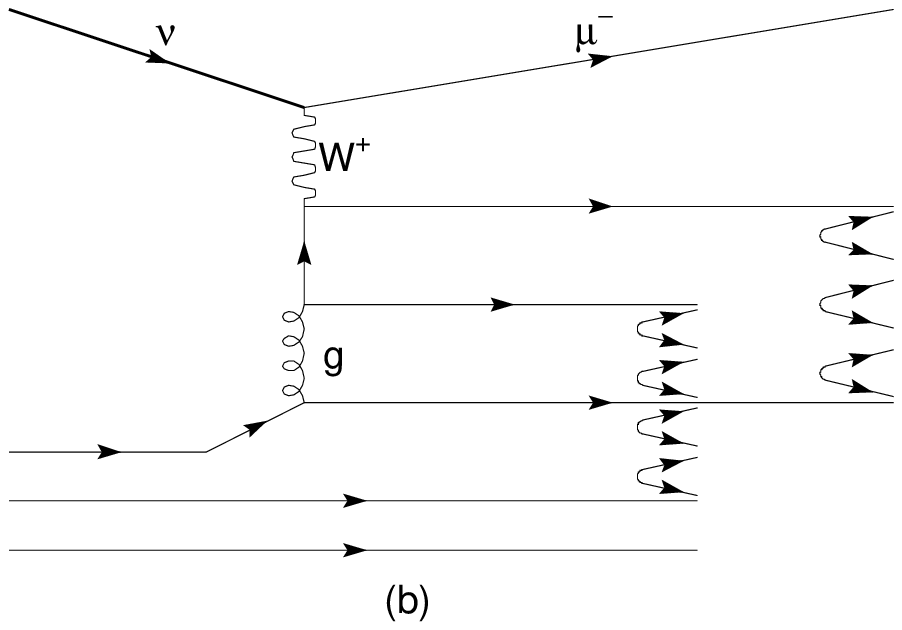
,angle=000,width=10.cm,height=8.0cm}}
\vspace*{.2in}
\caption{
Planar (a) and cylindrical (b) graphs contributing to the
reaction $\nu + A \rightarrow \mu + h + X$. Diagrams (a) and (b) 
describe processes in which the incoming neutrino interacts with a valence
quark or a sea $q\bar q$ pair, respectively. 
}
\label{graphs}
\end{figure}

The second ingredient entering the calculation of the spectrum
of eq.(\ref{def:lasp}) is the elementary semi-inclusive
spectrum. It can be calculated within the QGSM, based on the $1/N$ 
expansion, $N$ being the number of flavors or colors,  developed in \ci{kaid2}.
 According to ref.\ci{kaid1}, the elementary production process can be 
described in terms of planar and cylindrical graphs in the $s$-channel, as 
shown in fig. \ref{graphs} for the case of neutrino interactions
associated with the exchange of a $W$-boson. The planar graph of fig. 1 (a) 
describes neutrino scattering off a valence quark. According to
 \ci{diper,bfl4} its contribution to the process
 $\nu + N \rightarrow \mu + \pi + X$ reads
\beq
F_P^N(x_1,Q^2;z_1,p_{1t}) = z_1\phi_1(x_1,Q^2)
\left[ \frac{1}{3}D_{uu \rightarrow \pi} \left( \frac{z_1}{1-x_1},p_{1t}
 \right) + \frac{2}{3}D_{ud \rightarrow \pi} 
\left( \frac{z_1}{1-x_1},p_{1t} \right) \right]\ ,
\label{def:f1p}
\eeq
with
\beq
\phi_1(x_1,Q^2) = \frac{G^2mE}{\pi}\frac{x_1}{1-x_1}
\frac{m^2_W}{m^2_W+Q^2}d_v(x_1,Q^2)\ , 
\label{def:phi1}
\eeq
where $G$ is the Fermi coupling constant, $E$ is the energy of
the incoming $\nu$, $m_W$ is the $W$-boson mass and 
$x_1 \equiv x/y = Q^2/2(k \cdot p_\nu)$
is the Bjorken variable associated with a nucleon bound in a nucleus.
In eq.(\ref{def:f1p}), $d_v$ is the $d$-quark distribution inside a 
proton or a neutron, $D_{uu \rightarrow \pi}$ and $D_{ud \rightarrow \pi}$ 
are the fragmentation functions of $uu$ or $ud$ diquarks into positive or 
negative pions, $p_{1t}=p_t-(z/y)k_t$ is the transverse momentum of a pion 
produced on a nucleon carrying transverse momentum $k_t$ and $z_1=z/y$.

According to refs.\ci{kaid1,kaid2} the planar graph of fig.1 (a) corresponds
to one-Reggeon exchange in the $t$-channel, leading to the asymptotic
behavior $W_X^{\alpha_R(0)-1}$, where $W_X$ is the squared invariant
mass of the undetected debris and $\alpha_R(0)=1/2$ is the
Reggeon intercept. Hence, from
\beq
W_X=(k+q)^2=k^2+Q^2 \ \frac{ (1-{\widetilde x}_1) } { {\widetilde x}_1}\ ,
\label{invmass}
\eeq
where ${\widetilde x}_1=Q^2/2(k \cdot q)$, it follows that the Regge behavior 
of the planar graph of fig.1 (a)
at moderate and large $Q^2$ is given by
\beq
W_X^{-1/2} \sim \sqrt{ \frac{(1-{\widetilde x}_1)}
{{\widetilde x}_1 Q^2 } }\ .
\eeq
The fact that the $Q^2$-dependence of the graph of fig. 1 (a) is 
determined mainly by this Regge asymptotic is a consequence of the 
weak $Q^2$-dependence exhibited by the calculated $F_P^N$.

The second contribution to the spectrum, coming from the cylindrical graph of fig. 1 (b)
can also be obtained within the approach of ref.\ci{kaid2}. According to I, it can be 
written the form
\be
\nonumber
F_C^N(x_1,Q^2;z_1,p_{1t}) & = & z_1\ \phi_2(Q^2)\ \left[
L_1(x_1,Q^2;z_1,p_{1t}) + {\widetilde L}_1(x_1,Q^2;z_1,p_{1t}) \right. \\
& + &  \left. L_2(x_1,Q^2;z_1,p_{1t}) + {\widetilde L}_2(x_1,Q^2;z_1,p_{1t})
\right]\ , 
\label{def:fcn}
\ee
with
\beq
\phi_2(Q^2) = mE\ \frac{G^2}{\pi}\ \frac{m^2_W}{Q^2 + m^2_W}\ ,
\eeq
\be
\nonumber
L_1 & = & \int_{z_1}^{1-x_1} \frac{dy}{y}\ \int d^2k_t\ \left[ u_v(y,k_t;Q^2)\ 
D_{u\rightarrow\pi} \left(\frac{z_1}{y}, p_{1t}-\frac{z_1}{y}k_t \right) \right. \\
&  & \ \ \ \ \ \ \ \ \left. d_v(y,k_t;Q^2)D_{d\rightarrow\pi}
\left( \frac{z_1}{y}, p_{1t}-\frac{z_1}{y}k_t \right) \right]\ ,
\label{def:l1}
\ee
\be
\nonumber
L_2 & = & \int_{z_1}^{1-x_1} \frac{dy}{y}\ \int d^2k_t\ \left[ \frac{4}{3}f_{ud}(y,k_t;Q^2)
 D_{ud\rightarrow\pi} \left( \frac{z_1}{y}, p_{1t} - \frac{z_1}{y}k_t \right) \right. \\
\nonumber
&  &\ \ \ \ \ \ +  \left. \frac{1}{3} \left( f_{uu}(y,k_t;Q^2)D_{uu\rightarrow\pi} \left( \frac{z_1}{y},
 p_{1t}-\frac{z_1}{y}k_t \right) \right. \right. \\
&  & \ \ \ \ \ \ \ \ \ \ + \left. \left. f_{dd}(y,k_t;Q^2)
D_{dd\rightarrow\pi} \left( \frac{z_1}{y}, p_{1t}-\frac{z_1}{y}k_t \right) 
\right) \right]\ ,
\label{def:l2}
\ee
\beq
{\widetilde L}_1=d_{s}(x_1,Q^2)D_{u\rightarrow\pi}(z_1,p_{1t})
\label{def:tl1}
\eeq
and
\beq
{\widetilde L}_2=\int_{z_1}^{1-x_1}\frac{dy}{y}\ \int d^2k_t\ 
{\bar d}_{s}(y,k_t;Q^2)D_{{\bar d}\rightarrow\pi}
\left( \frac{z_1}{y}, p_{1t}-\frac{z_1}{y}k_t \right)\ .
\label{def:tl2}
\eeq
In the above equations $u_v$ is the 
distribution of the valence $u$-quark, $f_{uu}$, $f_{ud}$ and $f_{dd}$
are the distributions of $uu$-, $ud$- and $dd$-diquarks inside the
nucleon, $D_{u \rightarrow \pi},D_{d \rightarrow \pi}$ are the 
fragmentation functions of the $u$- and $d$-quark into a pion and 
$D_{dd \rightarrow \pi}$ is the fragmentation function of the $dd$-diquark
$dd$ into a pion. The explicit expression of the quark and diquark 
distributions and fragmentation functions, obtained within the approach 
of ref.\ci{kaid2}, are given in I.

According to refs.\ci{kaid1,kaid2} the diquark distribution
$f_{qq}(y)$ coincides with the distribution of the
corresponding valence quark evaluated at $1-y$, i.e. with $q_v(1-y)$, 
 if $q_v(y)$ is normalized to unity. The main contribution
to the pion spectrum is coming from target fragmentation,
i.e. the kinematical region corresponding to large values of $z_1=z/y$.
Within the approaches of refs.\ci{kaid1,kaid2} and \ci{dpm1,dpm2} 
the main contribution of sea quarks to the hadron spectrum
is associated with $n$-Pomeron exchange processes (with $n\geq 2$), which are
not taken into acount in this analysis as they provide a vanishingly small 
contribution at large $y$ \ci{kaid2,dpm1,dpm2}.
We only include sea quarks produced from gluon decay, as shown in fig. 1 (b).
In our approach the gluon is seen as a nucleon constituent, in addition to the three 
valence quarks. Therefore, we have used an average value of the gluon fraction
$<y_g>$ and normalized the valence quark distribution in a nucleon
to $1- \langle y_g \rangle$, the value $ \langle y_g \rangle \simeq 0.15 - 0.2$  
being taken from ref.\ci{gluon}. This procedure amounts to setting 
$f_{qq}(y)=q_v (1- \langle y_g \rangle - y)$.
Our calculations show that the contribution of fragmentation
of sea quarks and antiquarks (see the cylindrical graph of fig. 1 (b)),  
described by the terms $\widetilde {L}_1$ and $\widetilde {L}_2$, 
is much smaller than that coming from fragmentation of valence quarks
and diquarks, described by the terms $L_1$ and $L_2$ of eq.(\ref{def:fcn}). 

We assume a factorized form, similar to that of eq.(\ref{def:faykt}), for the 
quark and diquark distributions and fragmentation functions. The $k_t$ dependence 
of the quark distribution is again chosen of the Gaussian form 
\beq
g_q(k_t)=\frac{1}{ \langle k_{qt}^2 \rangle \pi}\ 
{\rm e}^{-k_t^2/ \langle k_{qt}^2 \rangle}\ ,
\eeq
where $ \langle k_{qt}^2 \rangle$ is the average value of the squared quark transverse
 momentum. For the fragmentation function we have used
\beq
g_{q(qq)\rightarrow h}(k_t)=\frac{\gamma}{\pi} \ {\rm e}^{-\gamma k_t^2}\ . 
\eeq
The cylindrical graph of fig. 1 (b) corresponds to one-Pomeron exchange in the $t$-channel, 
whose asymptotic behavior is $W_X^{\widetilde{\alpha}_P(0)-1}$, where
for the supercritical Pomeron \ci{kaid1,kaid2} the value of 
the exponent is given by $\Delta \equiv \widetilde{\alpha}_P(0)-1 \simeq 0.1-0.15$  
\ci{kaid3,kaid4}. Hence, the Pomeron asymptotic of the cylindrical graph of
fig. 1 (b) turns out to be $(k^2+Q^2(1-{\widetilde x}_1)/{\widetilde x}_1)^\Delta$.  
The $Q^2$-dependence of the cylindrical graph is dominated by the
supercritical Pomeron behavior, as $F_C^N$ of eq.(\ref{def:fcn}) depends 
weakly upon $Q^2$.

In conclusion, the relativistic invariant spectrum 
of pions produced in $\nu + N \rightarrow \mu + \pi + X$ processes
can be written in the form
\be
\nonumber
\rho_{\nu N \rightarrow \mu \pi X} \equiv  
\frac{zd^3\sigma}{dx_1dz_1dp_{1t}} & = & F_P^N(x_1,Q^2;z_1,p_{1t})
\left( \frac{(k^2+Q^2(1-{\widetilde x}_1)/{\widetilde x}_1)}{s_0} \right)^{-1/2} \\  
& + & F_C^N(x_1,Q^2;z_1,p_{1t})
\left( \frac{(k^2+Q^2(1-{\widetilde x}_1)/{\widetilde x}_1)}{s_o} \right)^{\Delta}\ ,
\label{def:rholn}                     
\ee
where $s_0 = 1$ (GeV/c)$^2$ is a parameter usually introduced in Regge
theory in order to get the correct dimensions.
Substituting $\rho_{\nu N \rightarrow \mu \pi X}$ of eq.(\ref{def:rholn}) and 
the nucleon distribution $f_A(y,k_t)$ into eq.(\ref{def:lasp}) one can calculate the
relativistic invariant spectrum of pions produced 
in the reaction $\nu + A \rightarrow \mu + \pi + X$.

\begin{figure}
\vspace*{1.in}
\centerline
{\epsfig{figure=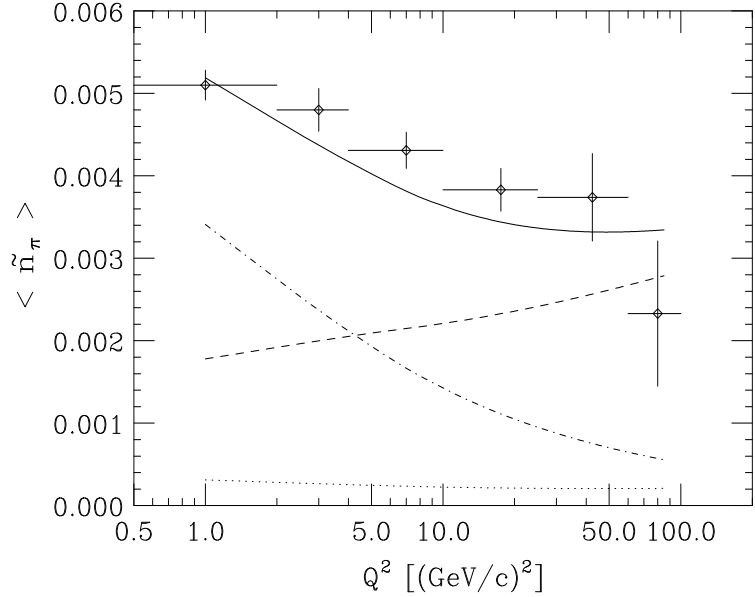
,angle=000,width=10.cm,height=8.0cm}}
\vspace*{.2in}
\caption{
$Q^2$-dependence of the normalized multiplicity 
of pions produced in the backward hemisphere in the
$\nu + C \rightarrow \mu^- + \pi^- + X$ reaction. The solid line shows
the results of the full calculation, including both graphs of 
fig. \protect\ref{graphs}. 
The dashed and dash-dot lines correspond to the separated contributions of the 
cylindrical graph of fig. 1 (b) and the planar graph of fig. 1 (a), respectively. 
The dots show the results obtained setting $P({\bf k},E) \equiv 0$ in the high
energy-momentum domain, not covered by the nonrelativistic calculation 
of ref.\protect\ci{bf1}. The experimental data are taken from 
ref.\protect\ci{NOMAD}.
}
\label{f2}
\end{figure}

The multiplicity of pions normalized to the cross section
of the process $\nu+A\rightarrow\mu+X$, $\sigma$, defined as 
\beq
\langle {\widetilde n}_\pi \rangle \equiv
\frac{ \langle n_\pi \rangle }{\sigma} = \frac{1}{\sigma}\ 
\int_{x_{min}}^{x_{max}} dx\ \int_{z_{min}}^{z_{max}} \frac{dz}{z}\ 
\int_0^{p_{\pi max}} d^2p_{\pi t}\ \rho_{\nu A \rightarrow \mu \pi X}\ ,
\label{def:npi}
\eeq
can also be split into two parts, corresponding to
the planar, or one-Reggeon exchange, diagram (fig. 1 (a)) and cylindrical, or
one-Pomeron exchange diagram (fig. 1 (b)). While at fixed ${\widetilde x}_1$ the first 
term decreases
as $Q^2$ increases, the second increases with a slope dictated by the 
value of $\Delta$. 
The $Q^2$-dependence of $\langle {\widetilde n}_\pi \rangle$
resulting from our approach is shown in fig. \ref{f2}, 
together with 
experimental data taken from ref.\ci{NOMAD}. 

The NOMAD collaboration carried out a study of inelastic $\nu$-$C$ 
interactions in which the negative pions emitted in the whole backward 
hemisphere, with respect to the incoming neutrino beam, were detected 
\ci{NOMAD}. The data of fig. \ref{f2} show the multiplicity 
of negative pions carrying momenta in the range $0.35 < p_\pi < 0.8$ GeV/c,
measured in a kinematical setup in which $W_X$ increases as $Q^2$ increases. 
Theoretical calculations have been performed applying the same kinematical 
conditions.

 The results obtained including 
both diagrams is represented by the solid line, while
the dashed and dash-dot lines correspond to the separated contributions of the 
cylindrical graph of fig. 1 (b) and the planar graph of fig. 1 (a), 
respectively.

The planar diagram provides the main contribution 
at small $Q^2$, while the cylindrcal one dominates at $Q^2 > 10$
(GeV/c)$^2$. The decrease of the planar graph contribution to 
$<{\tilde n}_\pi>$ and the increase of the cylindrical graph contribution 
with increasing $Q^2$ are both consequences of the 
kinematical setup of ref.\ci{NOMAD}, 
in which incresing $Q^2$ leads to an increase of the squared invariant 
mass $W_X$ (see eqs.(\ref{invmass}) and (\ref{def:rholn})). It clearly 
appears that 
both diagrams have to be included to explain the observed $Q^2$-dependence.
The fact that the solid line lies somewhat below the experimental data
at $Q^2 > 2$ (GeV/c)$^2$ is likely to be ascribed to the contribution of secondary
rescattering effects, which are not taken into account in our approach.

The dotted line has been obtained setting $P({\bf k},E) \equiv 0$ in the
domain of large energy and large momentum, not covered by the nonrelativistic 
calculations of ref.\protect\ci{bf1}. Comparison between the solid and dotted 
line shows that the dominant contribution to $\langle {\widetilde n}_\pi \rangle$ 
comes from the high momentum tail of the nucleon distribution. 

Numerical calculations have been carried out using values of the average squared
transverse momentum, entering eq.(\ref{def:ga}), in the range
$\langle k_t^2 \rangle = 0.12 - 0.14$ (GeV/c)$^2$. These values correspond to 
an average nuleon momentum $\sim$ 0.4 GeV/c, which seems to be reasonable
for processes dominated by the high momentum tail of the nucleon distribution. 
The associated ambiguity in the results is always within the experimental 
errors.

The other two parameters entering our calculations are the average quark and 
diquark transverse momentum and the slope of the Gaussian describing the
$k_t$-dependence of the  fragmentation functions. Their values 
have been taken from ref.\ci{ls}, where a modified QGSM, explicitly 
including the transverse motion of quarks and diquarks, has been developed.

\begin{figure}
\vspace*{1.in}
\centerline
{\epsfig{figure=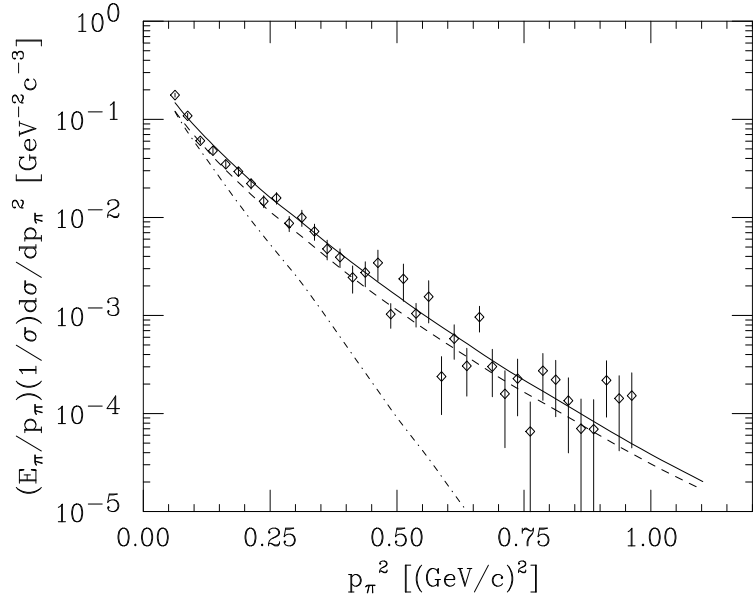
,angle=000,width=10.cm,height=8.0cm}}
\vspace*{.2in}
\caption{
$p^2$-dependence of the spectrum of backward pions produced in the 
reaction $\nu + C \rightarrow \mu^- + \pi^- + X$. The solid curve corresponds to the 
full calculation whereas the dashed line has been obtained including 
the cylindrical graph of fig. 1 (b) only. 
The dash-dot line shows the results obtained setting $P({\bf k},E) \equiv 0$ in the
high energy-momentum domain, not covered by the nonrelativistic calculation
of ref.\protect\ci{bf1}. The experimental data are taken from ref.\protect\ci{NOMAD}.
}
\label{f3}
\end{figure}

Besides the $Q^2$ dependence of the negative pion multiplicity, the 
NOMAD collaboration measured the semi-inclusive spectrum of negative pions, 
defined as  
\beq
\frac{2\pi}{\sigma} E_\pi \frac{d\sigma}{d^3p_\pi}
\equiv \frac{1}{\sigma}\ \frac{E_\pi}{p_\pi} \frac{d\sigma}{dp_\pi^2}\ ,
\eeq
in the kinematical region corresponding to pions emitted 
in the backward hemisphere with $p_\pi^2>0.05$ (GeV/c)$^2$ \ci{NOMAD}. 
In fig. \ref{f3} we compare the $p_\pi^2$-dependence of the experimental 
spectrum to that resulting from our approach.
 It clearly appears that the contribution 
of the cylindrical graph dominates the spectrum and that the inclusion of the 
high momentum tail of the nucleon distribution, corresponding to 
$p > 0.4$ GeV/c, is needed to describe the data.

In this paper we have analyzed inelastic neutrino-nucleus
 processes in which negative pions are emitted in the backward
emishpere. The main conclusions of our work, concerning both the reaction 
mechanism 
and role of nuclear structure, can be summarized as follows.
 The dominant contribution to the reaction comes
from target fragmentation, as shown by the calculation 
of the two topological graphs of fig. \ref{graphs}
 within the dual topological unitarization scheme. 
Two kinds of the elementary processes have to be
included: neutrino scattering off valence quarks, 
corresponding to one-Reggeon exchange in the $t$-channel
(planar graph, fig. 1 (a)), and neutrino scattering off sea quarks, 
corresponding to one-Pomeron exchange in the $t$-channel 
(cylindrical graph, fig. 1 (b)).
In the kinematical setup of ref.\ci{NOMAD} the contribution of the planar 
graph to the analyzed pion multiplicity
decreases as $\sqrt{1/Q^2}$ as $Q^2$ and $W_X$ increase, while
the contribution of the cylindrical graph increases as $Q^{2\Delta}$, 
 with $\Delta$ in the range $0.1-0.15$, as $Q^2$ and $W_X$ increase.  
The resulting $Q^2$-dependence of the the normalized multiplicity
of pions emitted in the backward hemisphere in the reaction
$\nu + C \rightarrow \mu^- + \pi^- + X$, turns out to be in fair agreement 
with the experimental data of ref.\ci{NOMAD}. 
As for the role of nuclear structure,  
 the dominant contribution is coming from the high momentum
tail of nucleon distribution, which can be described in terms of
overlaps of distributions of three-quark colorless objects \ci{bfl1}.
Including both graphs of fig. \ref{graphs}, our approach also provides a 
satisfactory description of the $p^2$-dependence of the measured 
pion spectrum.

We are indebted to A.B.Kaidalov and M.Veltri for many helpful discussions.
This work was partly supported by the U.S. Department of Energy. 
 

\clearpage

\clearpage

\end{document}